\documentstyle[aps,multicol,psfig,epsf,epsfig]{revtex}
\begin{document}
\def\pa{\parallel}
\def\pe{\bot}
\def\xx{\bf x}
\def\bea{\begin{eqnarray}}
\def\eea{\end{eqnarray}}
\def\be{\begin{equation}}
\def\ee{\end{equation}}
\let\a=\alpha \let\b=\beta  \let\c=\chi \let\d=\delta  \let\e=\varepsilon
\let\f=\varphi \let\g=\gamma \let\h=\eta \let\k=\kappa  \let\l=\lambda
\let\m=\mu   \let\n=\nu   \let\o=\omega    \let\p=\pi
\let\r=\varrho  \let\s=\sigma \let\t=\tau   \let\th=\vartheta
\let\y=\upsilon \let\x=\xi \let\z=\zeta
\let\D=\Delta \let\F=\Phi  \let\G=\Gamma  \let\L=\Lambda \let\Th=\Theta
\let\O=\Omega
\newcommand{\ie}{\hbox{\it i.e.\ }}         
\draft
\tightenlines
 
\title{Systems with superabsorbing states}
\author{
 Pablo I. Hurtado$^{*}$ and Miguel A. Mu\~noz$^{\dagger}$ }
\address{
Institute {\em Carlos I} for Theoretical and Computational Physics\\
and Departamento de E. y F{\'\i}sica de la Materia,\\
Universidad de Granada
 18071 Granada, Spain.\\    
}
\date{\today}

\maketitle
\begin{abstract}
 We report on some extensive analysis of a recently proposed
model [A. Lipowski Phys. Rev. E {\bf 60}, 6255 (1999)] with infinitely many 
absorbing states. By performing extensive Monte Carlo simulations
 we have determined critical exponents 
and show strong evidence that this model 
is not in the directed percolation universality class.
 The conjecture that
this two-dimensional model exhibits 
 a dimensional reduction
(behaving as one-dimensional directed percolation)
 is firmly disproven.
 The reason for the model not exhibiting 
 standard directed percolation scaling behavior is traced back
 to the existence of what we call
{\it superabsorbing sites},
 i.e. absorbing sites that cannot be directly activated
by the presence of neighboring activity in one or more than one directions.
 Supporting this claim we present two strong evidences:
(i) in one dimension, where superabsorbing sites do not appear at the
critical point, the system behaves as directed percolation, and (ii)
 in a modified two-dimensional variation of the model,
defined on a honeycomb lattice, 
for which superabsorbing sites are very rarely
observed, 
directed percolation behavior is recovered.
Finally, a parallel updating version of the model
exhibiting a nonequilibrium 
first-order transition is also reported.
\end{abstract}


\begin{multicols}{2}
\narrowtext

\section{Introduction}

Phase transitions separating active from  fluctuation-free
 absorbing phases appear in a vast group of physical phenomena
and models
as, for instance, directed percolation
\cite{reviews,Hinrichsen},  catalytic reactions \cite{catal}, 
 the pining of surfaces by disorder \cite{Barabasi},
 the contact process
\cite{CP}, damage spreading transitions \cite{damage}, 
nonequilibrium wetting \cite{firstorder2}, or  sandpiles
\cite{soc,soc2}. See \cite{reviews} and  \cite{Hinrichsen} for recent
reviews. 
Classifying these transitions into universality
 classes is a first
priority theoretical task. As conjectured by Janssen and Grassberger
\cite{conjecture} some
time ago and corroborated by a huge number of theoretical studies
 and computer simulations, systems exhibiting a continuous transition into a 
unique absorbing state with no extra symmetry or conservation law 
belong to one and the same universality class, namely that of 
directed percolation (DP). At a field theoretical level this class 
is represented by the Reggeon field theory (RFT) \cite{RFT}.

 This universality conjecture has been extended to include
multicomponent systems \cite{Lai}
 and systems with infinitely many absorbing states
\cite{Muchos,many}. On the other hand some other, 
less broad, universality classes
of systems with absorbing states
have been identified in recent years. They all 
include some extra symmetry or conservation law, 
foreign to the DP class.
 For example, if two symmetric absorbing states exist 
(which in many cases is equivalent to having activity parity-conservation 
 \cite{baw}), the universality class is other than DP, and the corresponding
 field theory differs from RFT  \cite{CT}.         
A second example is constituted by systems with absorbing states 
in which fluctuations occur only at the interfaces separating active 
from absorbing regions, but not in the bulk of compact active regions
(examples of this are the {\it voter model} or compact directed percolation
\cite{CDP}).
In this case the exponents are also non DP. A third and last example
is that of systems with many absorbing states
 in which the activity field is coupled
 to an extra conserved field.
This type of situation appears, for example, 
in conserved sandpile models, and 
has been recently shown to define a
new universality class \cite{soc,Conserved}.
   Apart from these and some few other well
 known examples \cite{other}, systems
with absorbing states belong generically into the DP universality class.

  Recently, Lipowski has proposed a very simple, biologically motivated 
model, exhibiting a continuous transition into an absorbing phase, and
claimed that this model shows a sort of ``superuniversality'', i.e.
in both 
one and two dimensions the model has the same critical exponents,
namely those of one-dimensional DP. Consequently, the 
system has been hypothesized to show a rather strange 
 ``dimensional reduction'' \cite{reduction} in two dimensions.
This conclusion, if confirmed, would break the Janssen-Grassberger 
conjecture, since it is not clear that any new symmetry or extra conservation
law is present in this model. In what follows we show what are the
physical reasons why this model does not show directed percolation
behavior: the presence of what we called {\it superabsorbing sites}
is at the basis of this anomalous behavior. We will discuss also how
DP can be restored by changing the geometry of the lattice
on which the model is defined. 

\section{The model}

 The Lipowski model is defined operationally as follows:
let's consider a square d-dimensional 
lattice. At a bond linking neighboring sites, $i, j$,
a random variable $w= w_{ij}$ is assigned.
 Different bonds are 
uncorrelated, and $w$ is distributed homogeneously in the interval 
 $[0,1]$.  At each site $i$ one defines $r_i$ as the
sum of the four bonds connecting this site to its four nearest neighbors.
If $r_i$ is larger that a certain threshold, $r$ (that acts as a control 
parameter) the site is declared active,
 otherwise the site is inactive
or absorbing.  Active sites are considered unstable; at each step
one of them is chosen randomly  and its
 four associated $w_{ij}$ bond variables
are replaced by four freshly chosen independent random values
 (extracted from the same 
 homogeneous probability distribution), and time is incremented by
an amount
$\Delta t = 1/n(t)$, where $n(t)$ is the number of active sites at that
time.  Critical exponents are defined as it is customary in the
realm of absorbing phase transitions \cite{reviews}.

   It is clear that for small values of $r$, for instance $r=0$, the
system will always be active, while for large enough values of $r$ 
 an absorbing configuration (with
$r_i < r$ for all sites $i$) will be eventually reached.
 Separating these two regimes we observe
 a critical value of $r$, $r_c$,
 signaling the presence of a continuous phase
transition.  In $d=1$ $r_c \approx 0.4409 $ \cite{Lip1}, while
for $d=2$ we find $r_c=1.38643(3)$.  As bond variables are continuous 
it is obvious that there is a continuous degeneracy of the absorbing
state (i.e. infinitely many absorbing configurations). 

 In the one dimensional case, all the measured critical exponents  
take the expected DP values \cite{Lip1}, compatible with theoretical
 predictions for 
systems with many absorbing states \cite{many,JSP}. The only
discrepancy comes from the fact that the spreading 
exponents $\eta$ and $\delta$ (see section IIIB for definitions)
 appear to be non-universal, but
the combination $\eta+\delta$ coincides with the DP expectation. This 
non-universality in the spreading  is however generic of one-dimensional 
systems with an infinite number of absorbing states \cite{PCP,JSP},
and therefore it does not invalidate the conclusion that the
system behaves as DP.

 In two dimensions the only measured critical exponent in \cite{Lip2}
 is the order parameter one,  $\beta$, which
 has been reported to take a value surprisingly close to the
one dimensional DP expectation, $\beta \approx 0.27$  \cite{Lip2}.
Based on this observation Lipowski claimed that
the system exhibits a sort of dimensional reduction. This possibility
would be very interesting from a theoretical point of view and
elucidating it constitutes the main original motivation of what
follows.         

Finally let us mention that for spreading experiments it was found that,
as happens generically in two-dimensional systems with many absorbing states
\cite{JSP,Chate}, the critical point is shifted, and
 its location depends on the nature of the absorbing
environment the initial seed spreads in.  In particular,
 the annular type of growth described in
\cite{Lip2} in the case of spreading into a favorable media is
typical of spreading in two-dimensional systems
 with many absorbing states, and it is well known 
to be described by dynamical percolation \cite{JSP,Chate}.

\section{Model Analysis}

In order to obtain reliable estimations for $\beta$ and 
 determine other exponents, we have performed extensive
Monte Carlo simulations in $d=2$ combined with finite size scaling analysis,
as well as properly defined spreading experiments.

\subsection{Finite size scaling analysis}
We have considered square lattice with linear dimension $L$ 
ranging
 from 32 to 256.  Averages are performed over a number of
independent runs
ranging from $10^2$ to $10^5$ depending on the distance to the critical point 
and on system size.
The first magnitude we measure is the averaged density
of active sites,
$ \rho(L,r,t)$, which for asymptotically large times converges to 
a stationary value $\rho(L,r)$. 
Observe that for small system
sizes the system always reaches an absorbing configuration in finite
time and therefore the only truly
 stationary state is $\rho=0$. 
In order to extrapolate the right asymptotic behavior in the active phase
one has to determine $\rho(L,r)$
averaged over the runs which have not reach an absorbing configuration.
A peculiarity of this system is that its convergence towards  a well
defined 
stationary state is very slow,  fluctuations around mean values
are extremely persistent 
and, therefore, a huge number of runs is needed in order 
to obtain smooth evolution curves. 
Owing to this fact, numerical studies are
rather costly from a computational point of view.
 The reasons underlying such anomalously long lived fluctuations
will be discussed in forthcoming sections.
The maximum times considered are $8 \times 10^5$ Monte Carlo steps
per spin; this is
 one order of magnitude 
larger than simulations presented in \cite{Lip2}.  
Near the critical point the
relaxation times are very large (larger than $10^5$) and, 
in order to compute stationary 
averages, transient effects have been cut off. We observe the 
presence 
of a continuous phase transition separating the active from 
the absorbing phase at a value of $r \approx 1.38$.

Assuming that finite size scaling holds \cite{FSS} in the vicinity of
the critical point point $r_c$, we expect
for values of $r < r_c$ (i.e. in the active phase)
\begin{equation}
\rho(L,r) \sim L^{-\beta/\nu_\perp} {\cal{G}}(L/\Delta^{-\nu_\perp})
\label{uno}
\end{equation}
where $\Delta=| r-r_c|$.
Right at the critical point, this corresponds to a straight 
line in a double logarithmic plot 
of $\rho(L,r)$ vs. $L$.
 
 In figure 1 it can be seen that, in fact, we
observe a straight line as a function
 of $log(L)$ for $r=1.38643(3)$ which
constitutes our best estimation of $r_c$. 
This finite size analysis
allows us to determine $r_c$  with much better precision 
 than in the previous estimations
\cite{Lip2}.  From the slope of the previous log-log plot we measure
$\beta/\nu_\perp = 0.57(2)$ which is quite far from both, the 
one-dimensional DP exponent
$\beta/\nu_\perp= 0.2520(1)$, and the two-dimensional value
$0.795(5)$. 

\begin{figure}
\centerline{
\epsfxsize=3in
\epsfysize=3in
\epsfbox{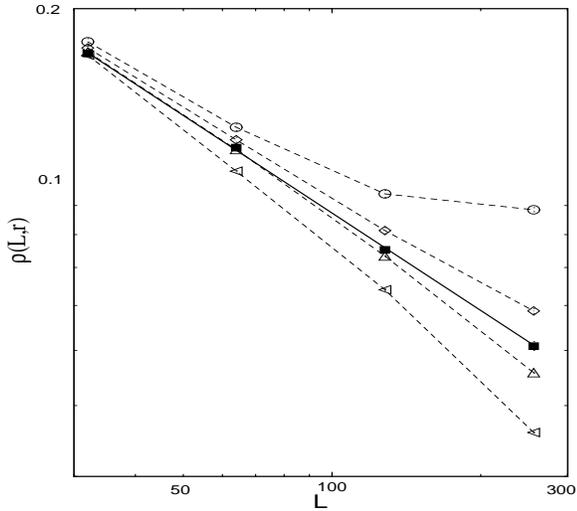}
}
\caption{
Density of active sites
as a function of $L$ (the linear system size) for 
different values of $r$: from top to bottom, $1.38630$,
$1.38640$, $1.38643$, $1.38645$, and $1.38650$ respectively.
The straight solid line corresponds to the critical point
$r_c=1.38643(3)$.}
\label{beta_nu}
\end{figure}        

We have  considered the larger available system size,
$L=256$, and studied the time decay of a fully
active initial state for values of $r$ close to $r_c$ in the 
active phase (see figure 2). The stationary values for
large values of $t$ should scale as  $\rho(L,r) \sim \Delta(L)^\beta$.
 From the best fit of our data (see figure 3) we determine both
 $r_c(L=256) \approx 1.38645 $ 
and $\beta =0.40(2)$.

 At the critical point, $\rho(r=r_c ,t) \sim t^{-\theta}$.
 From the asymptotic slope of the curve for $r_c(L=256)$
  in figure 2, we measure
$\theta = 0.275(15)$. In this way, we have already determined three 
independent exponents. 
>From these, using scaling laws, we can determine others, as for example
 $\nu_\perp= \beta/ (\beta/\nu_\perp) = 0.69(9)$ (to be compared with the 
DP prediction $1.09$ in $d=1$ and $0.733$ in two dimensions \cite{exp}).

  To further verify the consistency of our results we have  considered
$\rho(L,r)$  computed for different values of $r$ and $L$,
and assumed that $\rho(L,r) L^{\beta/\nu_\perp}$, 
depends on $r$ and $L$ through 
the combination $L^{1/\nu_\perp} \Delta $ \cite{reviews}. 
 In figure 4, we show the
corresponding data collapse 
which is  rather good when
the previously reported values of $\beta$ and $\nu_\perp$ are used.
In the inset
we verify that the data points are broadly scattered when one-dimensional 
DP exponent values are considered, 
showing that the dimensional reduction hypothesis
is not valid. Data collapse is neither observed
 using two-dimensional DP exponents; 
this provides a strong evidence that we are in the presence 
 of anomalous (non-DP) scaling behavior.
 Finally, let us remark that the observed scaling does
not extend over many decades for any of the computed steady state
magnitudes. Much better scaling is observed for spreading
exponents as will be shown in the following section.

\begin{figure}
\centerline{
\epsfxsize=3in
\epsfysize=3in  
\epsfbox{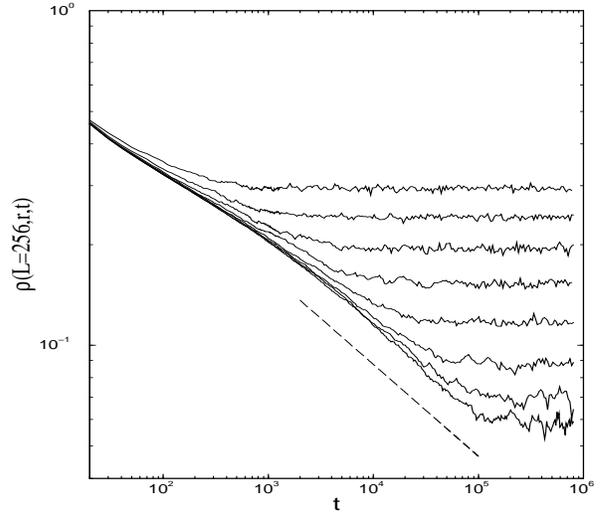}
}
\caption{
Time evolution of the density of active sites for $L=256$ and different 
values of $r$ in the active phase, namely, from top to bottom
$r=1.38143$, $1.38402$, $1.38527$, $1.38587$, $1.38616$, $1.38630$,
$1.38637$, and $1.38640$ respectively.
 From the slope of the straight dashed line 
we estimate $\theta=0.275(15)$.
}
\label{theta}
\end{figure}

 \subsection{Spreading experiments, and superabsorbing states }

In order to further verify and 
support our previous conclusion we have performed also
 spreading experiments as it is customarily done
 in systems with absorbing
states \cite{Torre,reviews}.
 These consist in locating a seed of activity at the center
of an otherwise absorbing configuration, and studying how it spreads 
 on average in that medium \cite{reviews}. In the absorbing 
phase the seed dies exponentially fast, propagates indefinitely in
the active phase, while the critical point corresponds to a 
marginal (power law) propagation regime \cite{reviews}. 

As said before, it is well established that two-dimensional
 systems with  
infinitely many absorbing states show some peculiarities
 in studies of the spreading of a localized activity seed.
 The absorbing environment surrounding the seed
can either favor or un-favor the propagation of activity depending on 
its nature  
 (see \cite{JSP,Chate} and references therein).
For the, so called, {\it natural} initial conditions \cite{reviews} the
critical point for spreading coincides with the bulk critical point, and 
standard DP exponents are expected. In order to generate such 
natural configurations one could start the system with 
some highly active configuration
and run the system right at the critical point;
 once it reaches an absorbing configuration
it can be taken as a natural or
 self-generated environment for spreading. 
An alternative, more efficient way of proceeding, inspired in sandpile
systems \cite{soc}, is as follows. 
 One considers an arbitrary absorbing configuration
 and runs a spreading experiment.
 Once the epidemic (or ``avalanche'' in the language of 
self-organized criticality

\begin{figure}
\centerline{
\epsfxsize=3in
\epsfysize=3in
\epsfbox{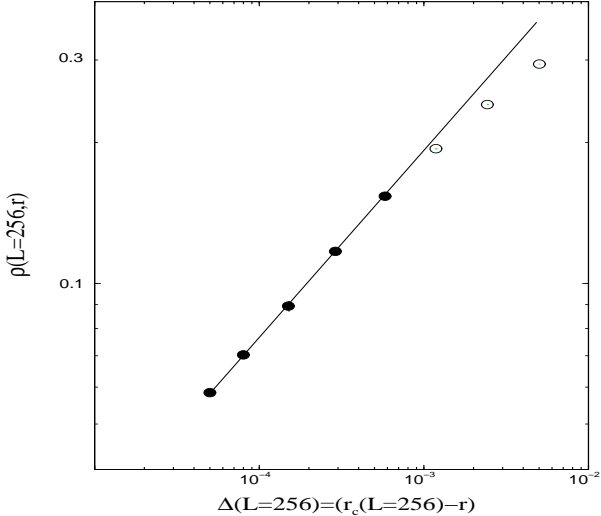}
}
\caption{
Stationary density of active sites as a function of the 
distance to the critical point, for $L=256$ and different values of
$r$ in the active phase (the same values reported in figure 2).
The best fit
gives $\beta=0.40(2)$ and $r_c(L=256) \approx 1.38645$. Filled (empty)
 circles are used to represent scaling (not scaling) points.
}
\label{beta}
\end{figure}

\noindent \cite{soc}) is over,
one considers the newly reached absorbing configuration as initial state
for a new spreading experiment avalanche.
 After iterating this process a number of times
the system reaches a statistically stationary absorbing state:
 the natural one
(see \cite{soc} and references therein).  Using this absorbing state for
 spreading leads to DP exponent values
 (and critical point) in systems with many absorbing states as 
for example the pair contact process \cite{PCP,dvz}.

 By following this procedure we have found a very peculiar
property of this model,
 that we believe to be at the basis of its deviating from DP.
If the initial seed is located for all avalanches in the same 
site (or small group of localized sites), 
as is usually the case, after a relatively 
small number of avalanches
 the system reaches an absorbing configuration such that it is
impossible to propagate activity
for any possible forthcoming avalanche
beyond a certain closed contour.
 For example, 
configurations as the one showed in figure 5a are generated.
The four sites at the center are the ones at which activity seeds 
are placed in order to start avalanches. White sites are active
and grey ones are absorbing. 
At  each marked-in-black site, the sum of
the three (black) bonds connecting it to sites other than a 
central one is smaller than $r_c-1=0.38643(3)$. In this way,
regardless the value of the bond connecting the site
to the central region the site remains inactive: it is a {\it superabsorbing
site}.  The existence of ``inactive forever'' sites have been 
already pointed out by Lipowski \cite{Lip2,Lip3}. 
 In the configuration showed in figure 5a 
activity cannot propagate out of the ``fence'' of superabsorbing
sites: the cluster of superabsorbing sites will remain frozen
indefinitely, and activity 

\begin{figure}
\centerline{
\epsfxsize=3in
\epsfysize=3in       
\epsfbox{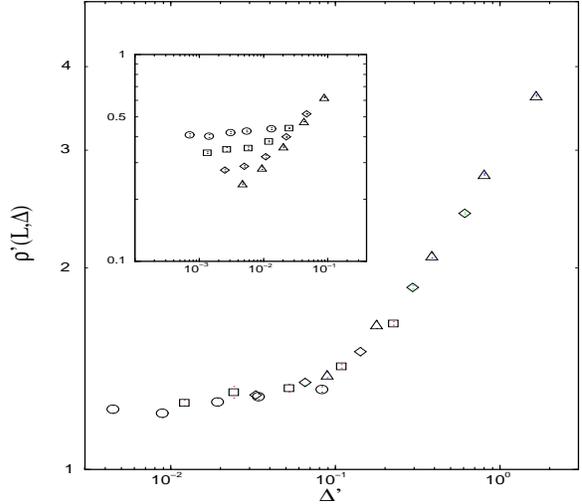}
}
\caption{
Data collapse for the density of active sites: $\rho' (L, \Delta) = 
\rho (L, \Delta) L^{\beta/\nu_\perp}$ and $\Delta' =\Delta L^{1/\nu_\perp}$. 
Using the obtained 
exponent values, $\beta/\nu_\perp \approx 0.57$, and $\nu_{\perp} \approx 0.69$,
 a reasonably good data collapse is observed.
In the inset we show an attempt to collapse data
using one-dimensional DP exponent values. There is no evidence of scaling
neither in this case nor using two-dimensional DP exponents.
}
\label{collapse}
\end{figure}

\noindent cannot possibly spread out. 
All avalanches will necessarily die after a few time steps. 
This type of blocking structure is quite generic,
 and appears in all experiments after some relatively short 
transient.

   In conclusion, this way of iterating spreading experiments leads always
to blocking closed configurations of superabsorbing sites
 instead of driving the system to
a natural absorbing configuration. 

  Observe that some activity put out of a blocking fence of sites
in figure 5a
could well affect any of the external bonds
 of the superabsorbing  sites (the dangling black bonds in figure 5a),
 converting the corresponding site 
to an absorbing or even an active one. 
Therefore, in order to overcome this  
difficulty of the frozen blocking 
configurations and be able to perform
 spreading experiments in some meaningful way,
we iterate avalanches by locating the initial seed 
at randomly chosen sites in the lattice. In this way there is always 
a nonvanishing probability of destroying 
blocking ``fences'' by breaking them from 
outside as previously discussed. Measurements of the different relevant
magnitudes are stopped when the system falls into an  
 absorbing configuration or alternatively whenever a linear
 distance $L/2$ from the avalanche origin is reached.
 Observe that in the second case
the dynamics has to be run farther
in order to reach a new absorbing configuration at which 
launching the next avalanche.

We monitor the following magnitudes:
 the total number of active sites in all the runs as a function of
time $N(t)$  (we also estimate $N_s(t)$ defined as the average 
number of active sites restricted to surviving runs),
 the surviving 

\begin{figure}
\centerline{
\epsfxsize=3in
\epsfysize=4in  
\epsfbox{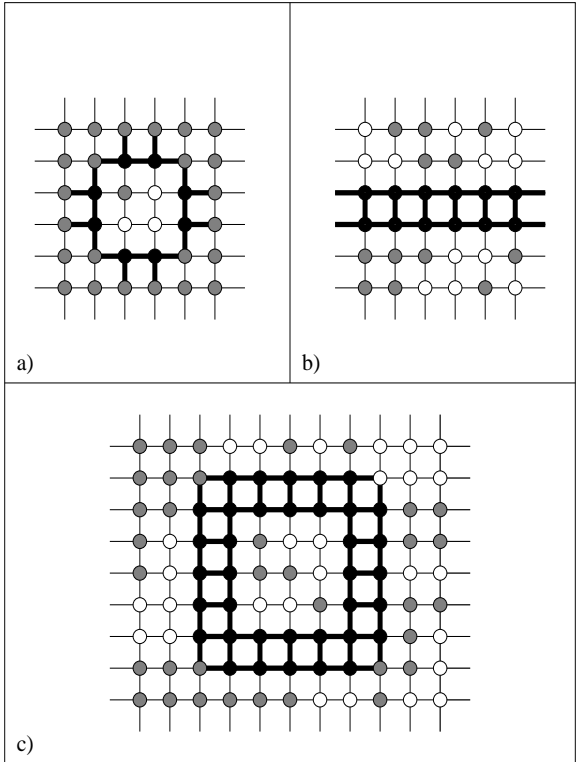}
}
\caption{Different frozen configurations of superabsorbing (black) sites. 
White (grey) color stands for active (absorbing) sites.
(a) Blocking configuration for spreading from the central cluster of
four sites. Black sites 
cannot change their state whatever the state or dynamics inside
the cluster might be. Black bonds remain also frozen.  
(b) Spanning frozen cluster of superabsorbing sites.
(c) Almost-frozen cluster of superabsorbing sites. This, and analogous
structures, can be destabilized from the outside corners.
}
\label{cluster1}
\end{figure}

\noindent probability  $P(t)$, and the average square 
distance from the origin, $R^2(t)$. At the critical point
these are expected to scale as $N(t) \sim t^\eta$, $P(t) \sim t^{-\delta}$
and $R^2(t) \sim t^z$. Results for this type of measurements
 are reported in figure 6.

 We obtain rather good algebraic behaviors at the previously
estimated critical point, $r_c$, confirming that the iteration-of-avalanches
procedure leads the system to a natural absorbing environment. Slightly
subcritical (supercritical) values of $r$ generate downward (upward)
 curvatures in this plot for all the four magnitudes.
Our best estimates for the exponents at criticality  
are: $z= 0.96(1)$, $\eta=0.05(1)$, $\delta=0.66(1)$ (see table 1).
To double check our results we also plot $N_s(t)$, which
 is expected to scale with an exponent 
$\eta +\delta$. An independent measurement of its slope in the log-log
plot gives $\eta+\delta=0.71(1)$, in perfect agreement with the
previously obtained results.

We can use these values to verify the hyperscaling relation \cite{Mendes,JSP}
\begin{equation}
\eta+\delta+\theta ={d z \over 2}.
\label{hyper}
\end{equation}
Substituting the found values for $z$ and $\eta+\delta$ we obtain 
$\theta \approx 0.25(2)$, compatible within error bars with the previously 
determined value $\theta =0.275(15)$. 
 
 One more check of the consistency of our 
results by using scaling laws is the following.
As $z=2 \nu_\perp / \nu_\pa$ \cite{exp},
 we can estimate $\nu_\pa$ from $z$ and
$\nu_\perp$. Then,  using $\nu_\pa$ 
and the fact that $\theta=\beta/\nu_\pa$ we obtain $\theta = 0.27(1)$, again
in excellent agreement with the directly measured value.

  In table 1, we present the collection of exponents and compare them
with DP values in both one and two dimensions \cite{exp}. There is no trace
of dimensional reduction: this model does not behave, at least up to the
scales we have analyzed, as any other known universality class.

\subsection{More about superabsorbing states}

 Let us recall our definition of superabsorbing states.
 A site, three of whose associated bonds take values
such that the sum of them is smaller that $r-1$, 
cannot be activated from the remaining direction by neighboring activity.
We say that this site is superabsorbing in that direction (or
it is in a superabsorbing state).  A site can
be superabsorbing in one or more than one directions.
Still a site in a superabsorbing state
can obviously be activated by neighboring
activity in any of the remaining directions (if any).
                                              
  Having stated the existence of frozen clusters in standard spreading
 experiments (when initialized from a  fixed localized set of sites),
one may wonder whether there are similar frozen structures in simulations
 started with an homogeneous initial distribution of activity, or in the
 modified type of spreading experiments we have just used (i.e. allowing
the initial seed to land at a randomly chosen site) in the neighborhood 
of the critical point. 

In principle, for any finite lattice, the answer to that question is 
affirmative.
In figure 5b we show the shape of a frozen cluster of superabsorbing
sites: any of the sites in it is superabsorbing with respect to the
corresponding outward direction, and it cannot be ``infected'' from
any of the other directions as neighboring sites are
 similarly superabsorbing.
If a cluster like that is formed (or put by hand on the initial state)
it will remain superabsorbing forever. However the probability to form 
such a perfectly regular chain is extremely small for large system sizes.
Observe also that in order to have a completely frozen two-site 
broad band structure it has to be unlimitedly
\begin{table}
\begin{center}
\begin{tabular}{|c|c|c|c|c|c|c|}
Model &  $\beta $ & $\beta/\nu_\perp$ & $\theta$ & $\eta$ & $ \delta$  & $ z$  \\
\hline
Lipowski &  $0.40(2)$  & $0.57(2)$  & $0.275(15)$  &$0.05(1)$&0.66(1)& $0.96(1)$   \\
\hline
 DP, $d=1$  &  $0.276 $   &  $0.252$   & $0.159$  &  $0.313$ & $0.159$   & $1.265$ \\
\hline
 DP, $d=2$  &  $0.583$    &  $0.795$   & $0.450$  &  $0.229$ & $0.450$   & $1.132$
\end{tabular}
\end{center}
\label{tabla}
\caption{Exponent values for the two dimensional Lipowski model
 and
directed percolation in both one and two dimensions. 
Figures in parenthesis denote
statistical uncertainty (note that error-bars are statistical errors
coming from power-law fittings, and therefore do not include eventual 
systematic corrections to scaling).
}
\end{table}

\begin{figure}
\centerline{
\epsfxsize=3in
\epsfysize=3in  
\epsfbox{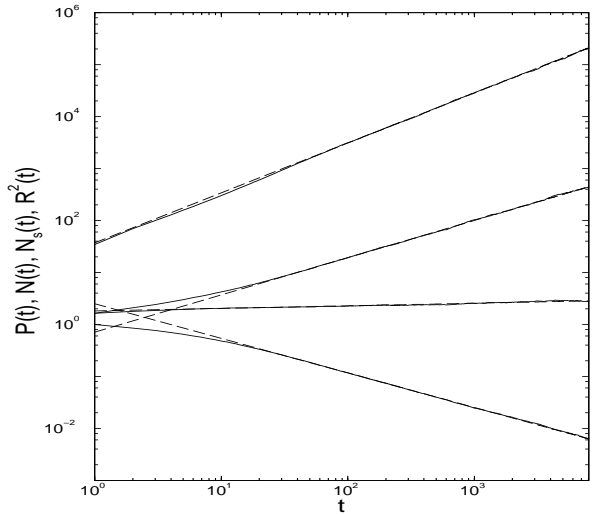}
}
\caption{Numerical results for spreading experiments. 
$R^2(t)$ (topmost curve), $N_s(t)$ 
 (second curve from above),
$N(t)$ (third curve from above),
and $P(t)$ (bottom curve).
>From the slopes we estimate
$z=0.96(1)$ and $\eta+\delta=0.71(1)$, $\eta=0.05(1)$ and $\delta=0.66(1)$
respectively.
}
\label{spreading}
\end{figure}

\noindent long (or closed if periodic
boundary conditions are employed). If instead it was finite,
 then sites at the corners
 would be linked to two external susceptible-to-change bonds and, therefore,
 themselves would be susceptible to become active: they would not be blocked
forever.
 In this way any finite structure of 
superabsorbing sites in the square lattice is unstable:
 it can be eaten up (though very slowly)
 by the dynamics, and is therefore not fully frozen.
   For instance, the cluster of superabsorbing 
sites represented in figure 5c is almost-frozen, but not really
frozen as it may lose its superabsorbing character from the 
outside corners
as previously described.  Analogously, any other cluster shape of 
superabsorbing sites may be destabilized from its outside corners.

  In conclusion, frozen clusters of superabsorbing sites do not appear
spontaneously. However, almost-frozen regions do appear and
may have extremely long life spans, specially 
 close to the critical point where activity is scarce, and therefore
the possibility of destabilizing them is small.  
In order to give an idea of how frequently superabsorbing sites appear
 we present in figure 7 a snapshot of a typical system-state
near the critical point.

 White corresponds to active sites, while the remaining sites are absorbing: 
in black we represent superabsorbing (in one or more than one directions)
 sites,
 while simple absorbing (not-superabsorbing) sites are marked in grey
color.
 Observe that superabsorbing sites are ubiquitous; in fact they percolate
 through the system.  Among them, about
one forth are superabsorbing in all four directions.
  
   Even though none of the clusters of superabsorbing sites is 
completely 
frozen, and in principle, activity could reach any lattice site, the 
dynamics is {\it glassy} \cite{glass} in some sense. For instance,
imagine an active region separated 
\begin{figure}
\centerline{
\epsfxsize=3in
\epsfysize=3in  
\epsfbox{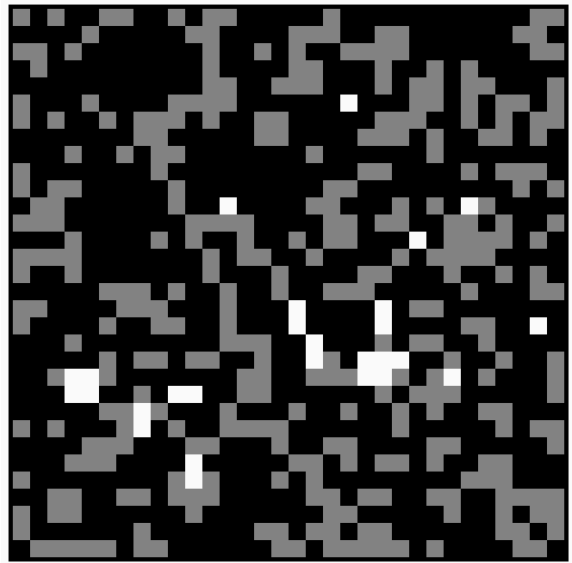}
}
\caption{Snapshot of a configuration in a $32 \times 32$
lattice in the stationary regime for a value of $r$ close to 
the critical point.
 White color denotes activity, black corresponds
to superabsorbing sites, while grey  stands for absorbing sites. 
Observe that superabsorbing sites percolate through the lattice.}
\label{snapshot}
\end{figure}
\noindent from an absorbing region by a line of
superabsorbing-in-the-direction-of-the-activity sites. In order to reach
the absorbing region, activity has to circumvent the superabsorbing barrier.
But near the critical point, where activity is scarce, barriers
of superabsorbing sites are intertwisted among them
 forming structures that,
even if not completely frozen, are very unlikely to be infected: activity
has to overcome them progressively in order to reach the interior
of superabsorbing regions.
This resembles  some aspects of
glassy systems for which degrees of freedom are hierarchically 
coupled and, at observable timescales, they may appear
 effectively frozen   \cite{glass}.    
 
 This phenomenology is certainly very different
from DP, and it is the reason why the relaxation towards
stationary states is so slow, and why deviations
from mean values are so persistent in numerical simulations.
 In particular, as superabsorbing regions are
long lived, the time required for the system to self-average is very 
large, and as near the critical point the probability of reaching an absorbing
state is large, in practice, 
the system does not have the time to self-average. 
Consequently, a huge amount of  
independent initial states and runs have to be considered in order
 to measure smooth  well behaved physical magnitudes \cite{SA}.
We strongly believe that this
 type of pathological dynamics is responsible 
for the departure of the Lipowski model from the DP universality class
in two dimensions.
  
 At this point one might wonder whether
the one-dimensional version of this model is essentially different.
Or in other words, why (one-dimensional) DP exponents are
 observed in $d=1$ \cite{Lip1}?.
 The answer to this
question is not difficult if one argues in terms of superabsorbing sites.
First of all notice that in $d=2$, $r_c > 1$. This means that just
by changing one bond, whatever the value of the output is, the site can stay
below threshold if the other three bonds sum less than $r_c-1$; this is
to say
superabsorbing states do exist at criticality.
  However in $d=1$,
$r_c=0.4409 < 1$. In this case by changing one bond value
it is always possible to activate the corresponding site:  superabsorbing
sites do not exist in $d=1$ at the critical point \cite{acla}. 
Once the ``disturbing'' ingredient is removed from the model, we
are back to the DP class
as general principles dictate.

\subsection{The honeycomb lattice}

 In order to further test our statement that superabsorbing states are
responsible for the anomalous scaling of the two-dimensional Lipowski model,
we have studied the following variation of it.
 We have considered the model defined
on a honeycomb lattice (with three bonds per site), and performed Monte Carlo
simulations.  In this case there is the (geometrical) 
possibility of having completely frozen
clusters of superabsorbing sites (see figure 8).
                                         
 The main geometrical difference from the
previous case comes from the fact that here cluster-corners are linked only
to one external bond, and therefore are more prone to form
frozen clusters.
 In principle, before performing any numerical analysis,
 there are two alternative possibilities:
either the critical point is located at a value of $r$ smaller than $1$ or
 larger than $1$.            
 In the first case, there would be no superabsorbing site (in analogy
with the one-dimensional case); in the second case pathologies associated 
with superabsorbing sites should be observed. 
The case $r_c=1$ would be marginal.
Finite size scaling analysis 
indicate the presence of a continuous phase transition
located at $r \approx 1.0092$ (very nearby the marginal case,
but significatively larger than $r=1$).

For Monte Carlo simulations, we have employed lattices
 of up to a maximum of
$256 \times 256 $ sites. All the observed phenomenology is
perfectly compatible with two-dimensional DP behavior. The dynamics
does  not show any of the
anomalies described for the square lattice case.  
 In particular, from the dependence of the stationary activity density
on system size  we evaluate  $\beta/\nu = 0.80(1)$;
 from the time decay at criticality
 $\theta=0.45(1)$, and finally $\beta=0.57(2)$;
 fully confirming consistency
with two-dimensional DP behavior. 
This result seems to be in contradiction with the two alternative 
possibilities presented above.
Let us now discuss why this is the case.  

   As the coordination number is $3$ in this case, 
the sum of two bond-values has to
be smaller than  $r_c -1 \approx 0.0092$ in order to have a superabsorbing 
site in the direction of the remaining bond at criticality.
As the
two bonds are independent random variables, the probability of creating a
superabsorbing site if the two of them are changed, is fewer
than $ 0.5 \%$, and the probability to generate frozen clusters 
(composed by six neighboring superabsorbing sites as shown in
 figure 8), is 
negligible at the critical point. In fact, we have not been able to observe
any of them in our simulations.
 This means that one should study extremely large system sizes
and extraordinarily  long simulations
in order to see  anomalies associated with superabsorbing sites,
otherwise, for any feasible simulation the behavior is expected to 
be DP-like. 
  The observation of DP exponents in this case strongly supports the 
hypothesis that superabsorbing states are at the basis of the 
anomalous behavior of the model on the square lattice.

 However, strictly speaking, the system should exhibit a (unobservable)
 first-order phase transition at $r=1$ in the thermodynamic limit.
Indeed, for values of $r$ larger than $1$ there is a finite,
 though extremely small, probability of   
creating frozen clusters of superabsorbing sites 
(as the one in figure 8). As
this is an irreversible process, 
after some (divergently long) transient there would be a
percolating network of frozen clusters of superabsorbing 
sites, and the only possible stationary state would be an absorbing one
with zero activity.
On the other hand, for values of $r$ smaller than unity, the probability of
creating superabsorbing sites is strictly zero, 
and there will be a nonvanishing density
of activity. As the density at $r=1$, almost independent of system size,
is $\rho  \approx 0.18$, the transition is expected to be discontinuous,
and therefore the DP transition observed in our simulations 
is merely a finite size effect, and should disappear for large
enough sizes and long times.
  In any case,
this first order transition is unobservable computationally. 

\section{Summary}

Summing up, we have shown that the two-dimensional 
Lipowski model does not belong to any known universality class.     
 We have measured different critical 
exponents by running Monte Carlo simulations started
 from homogeneous initial states and also by performing spreading
experiments.
 In any case, we find absolutely no trace of dimensional reduction,
 and there
is neither evidence for the system to behave as
two-dimensional DP. Instead, a novel scaling behavior
 is observed.      
The main relevant physical ingredient of this class is 
the presence of superabsorbing sites, and almost-frozen clusters
of superabsorbing sites which slow down enormously the dynamics.   

 The previous conclusion is strongly supported by two other observations: (i) 
the regular DP behavior observed in the one-dimensional version of the
model for which superabsorbing states do not appear at criticality, and 
(ii) the two-dimensional DP behavior observed for the two-dimensional model 
defined on a honeycomb lattice, for which the probability
of generating superabsorbing sites at criticality is almost negligible. 

   In general, superabsorbing sites can either arrange into completely
 frozen clusters or not depending on dimen-

\begin{figure}
\centerline{
\epsfxsize=3in
\epsfysize=3in  
\epsfbox{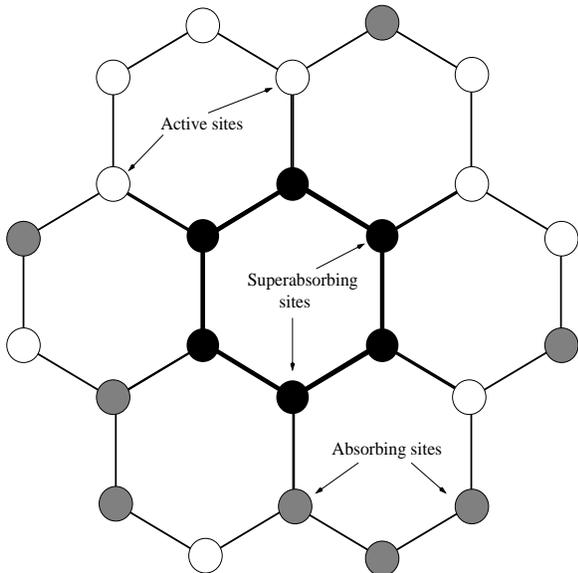}
}
\caption{Frozen cluster in the honeycomb lattice. This type
of frozen structure of superabsorbing sites remains indefinitely 
superabsorbing at the critical point. Black: superabsorbing sites. 
Grey: absorbing sites. White: active sites.
}
\label{cluster2}
\end{figure}

\noindent sionality, coordination
number and other system details. Let us distinguish three main cases:

(1) When completely 
frozen clusters of superabsorbing
sites appear below (or above) a certain value of the control parameter
but not above (below), first order
transitions are expected 
(as occurs in the multiplicative model discussed in appendix 2 
\cite{Lip3}).

(2)  If completely frozen clusters do not appear at criticality,
but instead almost-frozen clusters are present, 
we expect anomalous behavior (as 
occurs in the original Lipowski model \cite{Lip2}). 
   
(3) If neither frozen nor almost-frozen clusters are
 observed at criticality 
(as is the case for the one dimensional version of the model \cite{Lip1})
we expect standard directed percolation behavior.

  Two possible follow-ups of this work are: 

  (1) It would be worth studying in more realistic situations
as, for instance,
 in surface catalysis (dimer-dimer or dimer-trimer)  models
 \cite{Muchos}
whether effects similar to those described in this paper 
play any relevant role. In particular, for those models 
depending upon lattice and particle geometry there are
cases in which activity cannot propagate to neighboring regions,
but is constrained to evolve following certain directions or paths.
 It would be rather interesting to sort out whether
anomalies reported for those models \cite{Muchos} 
are related to the existence 
of superabsorbing states.
     
 (2) From a more theoretical point of view, an interesting question is:
 what is the field theory or Langevin 
equation capturing the previously described
 phase transition with superabsorbing states? and, how does it 
 change with respect to Reggeon field theory?.
Establishing what this theory looks like, would clarify greatly at a 
field theoretical level the effect of superabsorbing states on phase 

\begin{figure}
\centerline{
\epsfxsize=3in
\epsfysize=3in  
\epsfbox{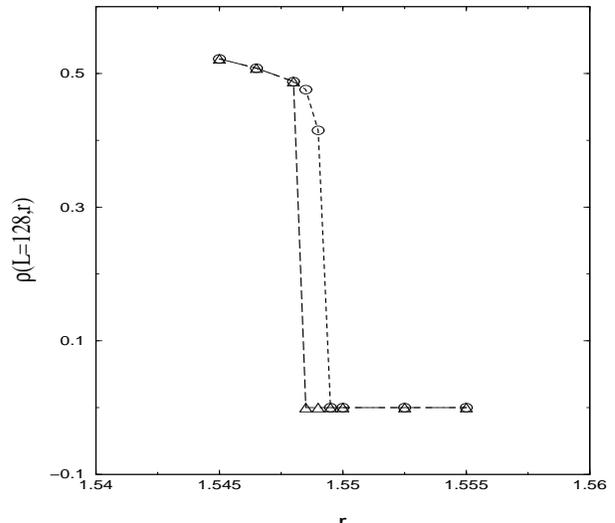}
}
\caption{Order parameter as a function of $r$ in the case
of parallel updating. The transition appears to be
 discontinuous in this case, exhibiting also 
a hysteresis loop.
}
\label{hysteresis}
\end{figure}

\noindent
transitions, and would permit to shed some light on 
the degree of universality of this anomalous phenomenology.
Our guess is that a
Reggeon field theory \cite{RFT,conjecture} 
with a spatio-temporal dependent anisotropic Laplacian term
(which, for example, would enhance, un-favor or forbid 
diffusion from certain sites in certain directions) could be a good
candidate to describe this new phenomenology.
  Analogously to what 
happens in field theoretical descriptions of other systems with many a
absorbing states \cite{Muchos,JSP}, the inhomogeneous Laplacian-term 
 coefficient should be described by a second physical field
 coupled to the activity field in such
a way that its fluctuations would vanish upon local
 absence of activity. 
 Further pursuing this
line of reasoning is beyond the scope of the present paper. 
 As long as this  program has not been completed, is not safe
to  conclude 
unambiguously that the anomalies described in this paper are
relevant in the limit of extremely large times and system sizes.

  \vspace{0.5cm}
{ \centerline{ \bf{  APPENDIX I}}}
\vspace{0.25cm}                          

 As an alternative attempt to speed up the dynamics, and examine further
some properties of the two-dimensional model, we have implemented
the microscopic dynamics replacing the original sequential updating by a
 synchronous or parallel one, i.e.
all active sites are ``deactivated'' simultaneously at each
Monte Carlo step, and all their associated bonds are replaced by
new random variables simultaneously. In this way, as random numbers do not
have to be extracted to sequentially select sites,
the dynamics is largely accelerated.
For this modified dynamics, we have examined some relatively
large system sizes, $L=256$, and concluded that the nature of the
transition is changed with respect to the sequential
updating case: in this case the transition is first order
and critical exponents cannot be defined. To show that this is the
case, in figure 9 we present the stationary activity curve.
The upper curve corresponds to simulations performed taking
an initial activity-density equal to unity. 
On the other hand, the lower curve is obtained by starting the system
with a natural absorbing configuration, and activating on the top 
of it a small percentage of sites (about a ten percent).

For values of $r$ in the interval $[\approx 1.545,
 \approx 1.555]$ the system reaches
different states depending upon the initial condition.
The presence of a hysteresis loop is a trait of the
transition first-order nature.
First order absorbing state transitions have been observed in other
contexts \cite{firstorder}.  However, we caution the reader that,
as the transition is found to occur at a value of $r$ for which the
probability of creating superabsorbing
sites is very large  (much larger than in the sequential case),
 and the dynamics is therefore extremely
anomalous and slow, it could be the case that the first order
character of the transition is only apparent. Extracting clean, conclusive
results in the critical zone is a computationally very expensive task, that
we have not pursued.

  \vspace{0.5cm}
{ \centerline{ \bf{  APPENDIX II}}}
\vspace{0.25cm} 
 Very recently, Lipowski has introduced a multiplicative version of his model
on the square lattice in which sites are declared active
 if the product of the four adjacent bonds
is smaller than a certain value of the control parameter $r$  \cite{Lip3}.
  Bonds take uncorrelated values in the interval $[-0.5, 0.5]$
extracted from a homogeneous distribution.
For values of $r$ smaller than $r=0$ 
there is a finite (not small) probability to
generate superabsorbing sites.
 In this case, it is not difficult to see that
isolated superabsorbing sites remain frozen forever.
 In analogy with the discussion of the honeycomb-lattice model,
a first order transition is expected at
 $r_c=0$ (as discussed also in \cite{Lip3}).
However, in this case, as the probability to create superabsorbing sites
is not negligible, the first order transition is actually observable.
  Based on a numerical measurement of $\beta$
 Lipowski concludes that the model
shares first-order properties with second-order features. 
In particular, 
the transition is clearly shown to be discontinuous,
 there is no diverging correlation length,
but $\beta$ is claimed to be however in the two-dimensional DP class.
 Our guess
is  that this apparent puzzle is simply due to a numerical 
coincidence, and that in fact there is no trait of any second-order phase
transition feature (observe that the fit
for beta in \cite{Lip3} spans for less than half a decade 
in the abscise of the log-log plot).

\vspace{0.5cm}
{ \centerline{ \bf{  ACKNOWLEDGMENTS}}}
 \vspace{0.25cm}               
We acknowledge P. L. Garrido and H. Hinrichsen
for useful discussions and criticisms,  R. Dickman 
for a critical reading of the manuscript,
 and valuable comments, and A. Lipowski and P. Grassberger 
for useful comments and for sharing with us valuable unpublished results. 
This work has been partially supported by the
European Network Contract ERBFMRXCT980183
 and by the Ministerio de Educaci\'on
under project DGESEIC, PB97-0842.      
\vspace{1em}

\noindent $^*$ electronic address: phurtado@onsager.ugr.es

\noindent $^\dagger$ electronic address: mamunoz@onsager.ugr.es

                                                 
\newpage
\end{multicols}

\begin{thebibliography}{99}
\bibitem{reduction}
  Dimensional reduction is not a new concept in statistical physics.
For example, quenched disordered magnetic systems were some time ago claimed
to behave in d dimensions as their corresponding pure counterparts in $d-2$
dimensions \cite{Parisi}. However, this results, is at odds with simple domain 
wall arguments, and
has recently proven to fail \cite{Brezin}.
\bibitem{acla} Observe that for values of the control parameter $r > 1$, well
into the absorbing phase, superabsorbing sites show up also in $d=1$; but
they do not affect the critical region.
\end{thebibliography}
\end{document}